\documentstyle[prl,twocolumn,aps,epsf]{revtex}

\begin{document}

\twocolumn[\hsize\textwidth\columnwidth\hsize\csname @twocolumnfalse\endcsname
\draft
\tolerance 500

\title{Interaction induced delocalisation for two particles in a
periodic potential}
\author{Julien Vidal$^{1}$,  Benoit Dou{\c c}ot$^{2}$, R\'emy Mosseri$^{3}$ and
Patrick Butaud$^{4}$}
\address{$^1$ Laboratoire de Physique des Solides, CNRS UMR 8502,
Universit\'e Paris Sud, B{\^a}timent 510, 91405 Orsay, France}

\address{$^2$ Laboratoire de Physique Th\'{e}orique et
Hautes \'Energies, CNRS UMR 7589, Universit\'{e}s Paris 6 et 7,\\
4, place Jussieu, 75252 Paris Cedex 05 France}

\address{$^3$ Groupe de Physique des Solides, CNRS UMR 7588,
Universit\'{e}s Paris 6 et 7,\\
2, place Jussieu, 75251 Paris Cedex 05 France}

\address{$^4$ Centre de Recherches sur les Tr\`es Basses
Temp\'eratures-CNRS associ\'e \`a l'Universit\'e Joseph Fourier\\
25, avenue des Martyrs, 38042 Grenoble, France}

\maketitle

\begin{abstract}
We consider two interacting particles evolving in a one-dimensional
periodic structure
embedded in a magnetic field. We show that the
strong localization induced by the magnetic field for particular
values of the flux
per unit cell is destroyed as soon as the particles interact.
We study the spectral and the dynamical aspects of
this transition.

\end{abstract}

\pacs{PACS numbers: 71.10.Fd, 71.23.An, 73.20.Jc}

\vskip2pc]

As shown by Anderson in
1958$^\protect{\cite{Anderson_localisation}}$, a quantum particle
in a disordered potential may be trapped in spatially localized eigenstates.
A natural question which arises is whether such a dramatic effect operates in
a many-body interacting system. This issue has been strongly revived
by a series of experiments
on $Si$ MOSFETs$^\protect{\cite{Kravchenko}}$ which seem to indicate
a metal-insulator
transition in two dimensions.
Since then, a lot of experimental and theoretical
activity has been dedicated to this highly controversial
topic$^\protect{\cite{Altshuler}}$.
Given the complexity of the full many-body systems, various groups
have studied the already
non trivial two-particle
problem$^\protect{\cite{Shepelyansky,Imry_TIP,Weinmann_TIP}}$.
In particular, D. L. Shepelyansky has shown convincingly that some
two-particle eigenstates
exhibit much larger localization length than single particle ones.
Note however that in
other situations such as in quasiperiodic systems, interactions may
generate strongly
localized two-particle eigenstates$^\protect{\cite{Barelli_TIP}}$. In
addition, it seems that
for repulsive interactions these interesting effects do not appear in
the vicinity of
the two-particle ground state, and this leaves open questions for the
many electron case.

Recently, it has been shown that an extreme localization mechanism
induced by the magnetic field
can lead to a complete confinement of the particle motion inside
Aharonov-Bohm cages$^\protect{\cite{Vidal_Cages}}$.
Contrary to the Anderson localization, this phenomenon occurs in pure
two-dimensional systems,  {\it i.e.}
without disorder, and is due to a subtle
interplay between the structure geometry and the magnetic field.
Two series of experiments have confirmed the existence of these
Aharonov-Bohm cages.
In the first one$^\protect{\cite{Abilio_T3}}$, superconducting wire
networks with the
adapted structure exhibit a striking reduction of the critical
current for the predicted
values of the magnetic field. The second
experiments$^\protect{\cite{Mailly_T3}}$ measure
the magneto-resistance oscillations in two-dimensional mesoscopic
structures with a small number
of conduction channels and large electronic mean free path. A clearcut dip at
half a flux quantum per loop is observed and confirms the presence of this
peculiar localization process.

It is therefore very natural to ask whether these cages survive for
interacting particles.
In this letter, we present an exacly solvable model for two
interacting particles under a
magnetic field. To simplify, we deal with a quasi-one-dimensional
model which exhibits
Aharonov-Bohm cages. We show that for half a flux quantum per loop,
dispersive two-particle bound states appear even for repulsive local
interaction. In this system,
the two-particle ground state is non dispersive but the first dispersive band
is rather close in energy.
Slightly away from these remarkable fluxes, these bound states survive
until they merge in a two-particle continuum. Finally, we are led to
speculate that a finite
repulsive local interaction is able to turn the fully localized non
interacting system into a
strongly correlated metal provided the electron density is large enough.\\

\begin{figure}
\vspace{-2mm}
\centerline{\epsfxsize=135mm
\epsffile{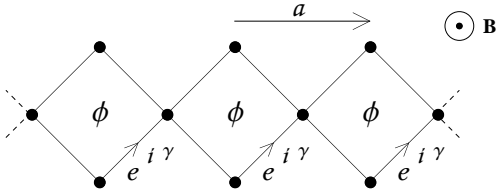}}
\vspace{-30mm}
\caption{\label{chap}
Square chain under a uniform magnetic field.The magnetic
flux per unit cell is denoted by $\phi$ and $\gamma=2 \pi \phi / \phi_0$.}
\end{figure}

We consider a one-dimensional chain of square loops
  with periodic boundary conditions displayed in Fig.~\ref{chap}
embedded in a uniform
perpendicular magnetic field ${\bf {B}}$, which is a bipartite
periodic structure with three
sites per unit cell.
As we shall see, the various characteristics of this system are similar to
those discussed in Ref. \cite{Vidal_Cages} for two-dimensional tilings.
Hereafter, we fix the total polarization to zero which
is equivalent to consider two particles with opposite spin
($\uparrow$ and $\downarrow$).

Let us consider the standard Hubbard hamiltonian~:
%
%
\begin{equation}
H=\sum_{\langle i,j \rangle, \sigma=\uparrow,\downarrow}
t_{ij} \: c^\dagger_{i,\sigma}\,c_{j,\sigma}+
U \sum_{i} n_{i,\uparrow} \, n_{i,\downarrow}
\mbox{,}
\label{hamil}
\end{equation}
%
%
where $c^\dagger_{i,\sigma}$ (resp. $c_{i,\sigma}$) denotes the creation (resp.
annihilation) operator of a fermion with spin $\sigma$,
$n_{i,\sigma}=c^\dagger_{i,\sigma}\,c_{i,\sigma}$ the  density of
spin $\sigma$ fermion on site $i$,
and
$\langle \ldots \rangle$ stands for nearest neighbor pairs.
Note that, since the particle considered here are fermions, the
interaction term $U$ is only
efficient in the singlet sector where the orbital part of the wave
function is symmetric.
When ${\bf B}=0$, the hopping term $t_{ij}=1$ if $i$ and $j$ are nearest
neighbors and $0$ otherwise. In the presence of a magnetic
field$^\protect{\cite{Peierls}}$,
  $t_{ij}$ is multiplied by a phase factor $e^{i \gamma _{ij}}$ involving the
vector potential ${\bf {A}}$~:
%
%
\begin{equation}
\gamma _{ij}={\frac{2\pi }{\phi _{0}}}\int_{i}^{j}{\bf A}.\mbox{d}{\bf l}
\mbox{,}
\end{equation}
%
%
where $\phi _{0}=hc/e$ is the flux quantum. For convenience, we choose a gauge
in which only one hopping term per unit cell is modified (see in
Fig.~\ref{chap}).
The whole spectrum only depends on the reduced flux $f=\phi
/\phi _{0}$ where $\phi =B a^{2}/2$ is the magnetic flux through an
elementary square ($a$ is the unit cell vector length).
%
%
%
%
%
Let us first analyze the one-particle problem. In this case, the
translation invariance
of the system along the chain direction allows one to
straightforwardly compute the
one-particle spectrum that consists of three bands~:
%
%
\begin{equation}
\varepsilon_{\alpha}(k)= 2 \alpha \:\sqrt{1+\cos (\gamma/ 2) \: \cos(k a)}
\hspace{5mm} k\in [0,2\pi/a]
\mbox {,}
\end{equation}
%
%
where $\alpha=0,\pm 1$ is the band index and $\gamma=2 \pi f$.
The weight of each band in the normalized density of states equals $1/3$.
The existence of a non dispersive band at $\varepsilon=0$ is simply due
to the bipartite character of the structure, its degeneracy being equal to
the difference between the number of sites of each family.
The most striking feature is that for $f=1/2$ (half a flux quantum
per unit square),
the spectrum is made up of three non dispersive bands. As discussed
in Ref. \cite{Vidal_Cages},
this property leads to a complete {\em lock-in} of any wave packet spreading
inside the so-called Aharonov-Bohm cages. One thus
has a transition induced by the magnetic field.

For $U=0$,
the two-particle spectrum is the addition of the one-particle spectra
so that the
eigenenergies are labelled by four quantum numbers~:
%
%
\begin{equation}
\varepsilon_{\alpha_\uparrow,\alpha_\downarrow}(k_\uparrow,k_\downarrow)=
\varepsilon_{\alpha_\uparrow}(k_\uparrow)+\varepsilon_{\alpha_\downarrow}(k_\downarrow)
\mbox{,}
\end{equation}
%
%
where $\alpha_{\sigma}=0,\pm 1$ (resp. $k_\sigma$) is the band index
(resp. the wave vector)
of the spin $\sigma$ particle. Thus, for $f=1/2$, the spectrum
consists of five non dispersive
bands corresponding to $\varepsilon=0,\pm2 ,\pm 4 $ and the space
evolution of any two-particle
wave function is confined in an Aharonov-Bohm
cage that is merely the superposition of each one-particle cage.

\begin{figure}[h]
\vspace{-30mm}
\centerline{\epsfxsize=120mm
\epsffile{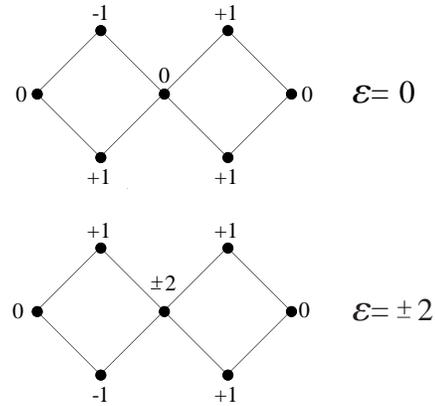}}
\vspace{-88mm}
\caption{\label{eigen}
Eigenstates of the one-particle problem for $f=1/2$ (non normalized
cage solutions).}
\end{figure}

We now address the interacting case where $U\ne 0$. The main question
is whether or not the latter system
remains frozen when the particles are interacting. In other words,
can a (local) interaction term
destroy the cages and autorize any propagation~?
In general, a two-particle problem with on-site interaction in a
$D$-dimensional structure can be
viewed as a single particle one in a $2D$-dimensional structure with
a local potential in the
hyperplane corresponding to a double occupancy of a site. Taking
advantage of the translation
invariance, this problem can then be mapped onto a (continuous)
family of $D$-dimensional problems
with a finite number of impurity sites which are often easier to
solve$^\protect{\cite{Caffarel_Mosseri}}$. The same approach could in principle
be used here but given the
very special nature of the non interacting system for $f=1/2$, it is
most easily carried out
by using the minimally extended one-particle eigenstates displayed in
Fig.~\ref{eigen}.
%
%
%
%
%
These states  have non vanishing amplitude only on a finite number of
sites and thus reflect the
absence of  propagation at $f=1/2$.
We denote them by $|i,\varepsilon_\alpha \rangle$,
where $i$ is the cell number centered on each 4-fold coordinated site and
$\varepsilon_\alpha=0,\pm 2$.

\begin{figure}
\centerline{\epsfxsize=100mm
\epsffile{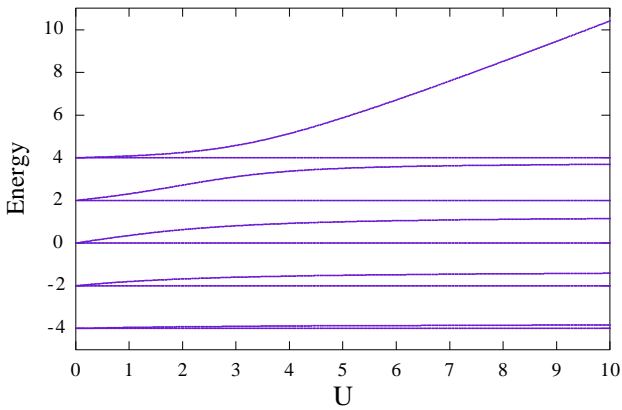}}
\vspace{-5mm}
\caption{\label{nondisp}
Non dispersive part of the two-particle spectrum at $f=1/2$ {\it
versus} the interaction.}
\end{figure}

The two-particle state space is then given by the tensor
product of the one-particle state~:
%
%
\begin{equation}
|i,\alpha; i',\alpha' \rangle=
|i,\varepsilon_{\alpha} \rangle_\uparrow \otimes
|i',\varepsilon_{\alpha'} \rangle_\downarrow
\mbox{,}
\end{equation}
%
%
for all $i,i',\alpha,\alpha'$.
It is worth stressing that since the $|i,\varepsilon_{\alpha} \rangle$ are
confined eigenstates of the one-particle hamiltonian, most of these
two-particle eigenstates
are not affected by $U$.
This local interaction term only acts on states for which the two
particles have a non vanishing probability to be
on the same site, {\it i.e.} such that $|i-i'| \leq 1$.
As a result, the number of states which are sensitive to $U$ scales
linearly with
the total number of cells $N$ whereas in the generic case
($f\neq1/2$), it is proportional
to $N^2$. To proceed further, let us remark that the  states
sensitive to $U$ are space-symmetric~:
%
%
\begin{equation}
|i,{\alpha}; i',{\alpha'} \rangle_S=
{1 \over \sqrt 2} \left(|i,{\alpha}; i',{\alpha'} \rangle+
|i',{\alpha'}; i,{\alpha} \rangle \right)
\mbox{,}
\end{equation}
%
%
if $i \neq i'$ or $\alpha \neq \alpha'$, and
$|i,{\alpha}; i,{\alpha} \rangle_S=|i,{\alpha}; i,{\alpha} \rangle$.
Moreover, since the problem is invariant under a translation of the
center of mass,
it is convenient to build Bloch waves~:
%
%
\begin{eqnarray}
|\varphi_{0}(\alpha,\alpha',K) \rangle&=& {1 \over \sqrt{N}} \sum_{n=0}^{N-1}
e^{iK n a} |n,{\alpha}; n,{\alpha'} \rangle_S\\
|\varphi_{1}(\alpha,\alpha',K) \rangle&=& {1 \over \sqrt{N}} \sum_{n=0}^{N-1}
e^{iK n a} |n,{\alpha}; n+1,{\alpha'} \rangle_S
\mbox{,}
\end{eqnarray}
%
%
where $K=2\pi j / N a \:\:\: (j=0, N-1)$ is a wave vector lying in
the first Brillouin zone.
In the following, we shall always considerthe limit where $N$ tends
to infini\-ty.
One then has to calculate the matrix element of the hamiltonian in
each irreducible representation
labelled by $K$, and restrict this analysis to the $(15 \times 15)$ subspace
generated by $|\varphi_{0}\rangle$ and $|\varphi_{1}\rangle$.
For any $(\alpha,\alpha',\beta,\beta',K_l,K_m)$ one has~:
%
%
\begin{eqnarray}
\langle \varphi_{0}(\alpha,\alpha',K_l)
|H|\varphi_{1}(\beta,\beta',K_m) \rangle
&=& 0\\
&& \nonumber\\
\langle \varphi_{1}(\alpha,\alpha',K_l)
|H|\varphi_{1}(\beta,\beta',K_m) \rangle
&=& \lambda_{\alpha,\alpha',\beta,\beta'} \: \delta_{l,m}
\mbox{,}
\end{eqnarray}
%
%
where $\lambda_{\alpha,\alpha',\beta,\beta'}$ is a
$K_{l,m}$-independent scalar and
where $\delta_{l,m}$ is the usual Kronecker symbol. This implies that
the eigenvalues
of the $(9\times 9)$ subspace generated by the $|\varphi_{1}\rangle$
are non dispersive.
These eigenvalues are given by the five $U$-dependent roots of the
characteristic polynomial~:
%
%
\begin{equation}
P(\varepsilon,U)=\varepsilon^5-U \varepsilon^4- 20 \,
\varepsilon^3+16U\varepsilon^2+64 \varepsilon
-24 U
\mbox{,}
\end{equation}
%
%
and four $U$-independent values $\pm2,0$ (two-fold degene\-rated) resulting
of additional symmetry between $\alpha$ and
$\alpha{'}$. The non dispersive part of the
spectrum$^\protect{\cite{additional}}$ is shown in
Fig.~\ref{nondisp} for $U\geq 0$.
%
%
%
%
%
We emphasize that since the spectrum of the Hubbard hamiltonian
(in a bipartite structure) is odd under the transformation
$U\rightarrow -U$$^\protect{\cite{Sutherland_Bethe}}$,
we can restrict our analysis to the repulsive case.
In the large $U$ limit, the five roots of $P$ tends toward $\pm
\sqrt{ 8\pm 2 \sqrt{10}},U$~;
this latter value simply corresponding to a situation where the two
particles are localized
on the same site (anti-bonding state).

\begin{figure}
\centerline{\epsfxsize=100mm
\epsffile{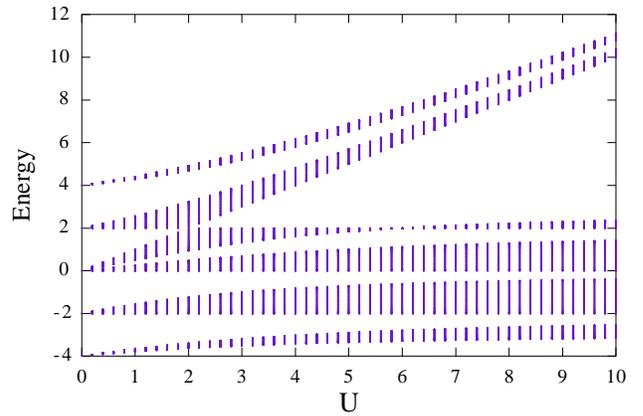}}
\vspace{-5mm}
\caption{\label {disp}
Dispersive part of the two-particle spectrum at $f=1/2$ {\it versus}
the interaction.}
\end{figure}

A much more interesting component of the spectrum is provided in the
$(6\times 6)$
subspace$^\protect{\cite{rem2}}$  generated by the $|\varphi_{0}\rangle$.
In this subspace, the eigenvalues
are given by the roots of the following characteristic polynomial~:
%
%
\begin{eqnarray}
Q(\varepsilon,U,K)&=&\varepsilon^6-2 U \varepsilon^5+(U^2-20) \, \varepsilon^4+
28 U\varepsilon^3 +\nonumber\\
&&8\left(8-U^2\right)\varepsilon^2-
4U(14-3\cos(K a)) \varepsilon+ \nonumber \\
&&4U^2(2+\cos(K a))
\mbox{.}
\end{eqnarray}
%
%
Contrary to the previous case, these eigenvalues obviously depend on
$K$, and the associated
non degenerated eigenstates are extended (Bloch-like). This
dispersive component of the spectrum is
displayed in Fig.~\ref{disp}.
%
%
%
%
%
In the large $U$ limit, the asymptotic eigenvalues are given by
$U$ (twice degenerated) and $\pm \sqrt{ 4\pm 2 \sqrt{2-\cos(K a)}}$, and
there still remains a $K$-depen\-dent component in the spectrum.

The physical consequences
of this dispersion are important. Indeed, let us consider a generic
two-particle initial state having
a non zero overlap with one of the
$|\varphi_{0}\rangle$$^\protect{\cite{rem1}}$.
The emergence of dispersive states for $U\neq 0$ indicates that it is
now possible for this wave packet
to spread over the whole system whereas it was completely trapped
inside the Aharonov-Bohm cage in the
non interacting case.
Moreover, since the dispersive eigenstates are extended, the
propagation is ballistic.

\begin{figure}
\centerline{\epsfxsize=85mm
\epsffile{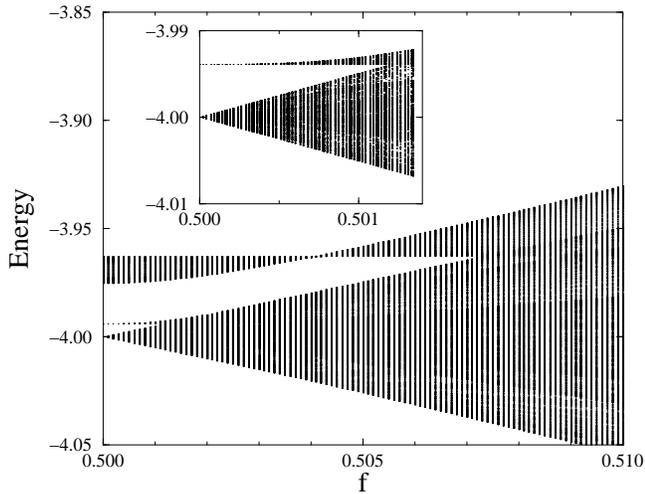}}
\caption{\label{butterflyTIP}
Low-energy spectrum for $U=0.01$ as a function of the reduced flux.}
\end{figure}

For other values of the reduced flux, the full solution of the
two-particle problem cannot be
cast in a simple analytic form, although it is possible to reduce it
into a scattering
problem for one particle moving on a chain in the presence of three
static impurities.
Nevertheless, let us analyze the neighborhood of the half-flux parametrized by
$\delta f=|f-1/2|$.
For $f\neq 1/2$, all the single-particle eigenstates become extended,
except those corresponding
to the flat band at $\varepsilon=0$.
The non dispersive two-particle states which are insensitive to $U$ at $f=1/2$
evolve into several two-particle continua, whose band width scales as
$\delta f$ for small
$\delta f$.
The non dispersive states which are sensitive to $U$ become dispersive with
a band width scaling as $(\delta f)^2/U$ (see inset in
Fig.~\ref{butterflyTIP}).
The corresponding wave functions still exhibit a binding of the two particles.
Finally, as displayed in Fig.~\ref{butterflyTIP}, the dispersive
states at $f=1/2$ remain dispersive.
All
these dispersive bound states evolve smoothly as a function of
$\delta f$ until their energies
merge in the two-particle continuum, which occurs for $\delta f \sim
U$ at small $U$.
As $\delta f$ further increases, the total number of bound states
gradually decreases.
%
%
%
%
%

Since, in this study, the effect of the interaction term between the
two particles is clearly to
induce a delocalization process, one can expect that for finite
density of particles, a subtle
correlated conducting state will emerge. Note however that in this
system, the energy
of two particles is minimal when they do not interact, either because
they are far
apart or because their orbital wave function is antisymmetric.
We therefore conjecture that the
many body ground state will remain localized up to a filling factor
equal to 1/3. At this point,
the lowest flat band is completely filled with a fully polarized electron sea.
The next electron
will have an opposite spin and is thus likely to delocalize along the
chain. This problem obviously
deserves futher investigations.
Let us also remark that, here, we consider a quasi-one-dimensional model
to simplify the calculations, but
a similar physics is expected for other tight-binding models which
exhibit single-particle confinement
inside Aharonov-Bohm cages. Nevertheless, in more general systems, we
do not expect to see
non dispersive states sensitive to $U$. In this context, an
interesting open question is
whether it is possible to find
situations where the ground state of the two-particle spectrum is dispersive,
even for repulsive interaction.

How could this interaction induced delocalization be obser\-ved
experimentally? The two-particle
system might
be accesible in a Josephson junction or a quantum dot array in the
Coulomb blockade
regime. Such experimental situations shall be described by a model
with on-site disorder,
but, very likely, some two-particle states would still exhibit a much
larger localization length
than single particle ones.
Ongoing experiments on ballistic semi-conducting networks with
special two-dimensional
geometry$^\protect{\cite{Mailly_T3}}$ may also manifest
some interaction effects on this Aharonov-Bohm localization. In these
structures, the interaction
strength can be varied by changing the electronic density or by
polarizing the system with a
tilted magnetic field. Clearly, a better understanding of the many
electron case needs to be
achieved.\\

We would like to thank Cl. Aslangul, B. Delamotte, D. Mouhanna and
M. Tissier for fruitful discussions.


\begin{thebibliography}{10}
\bibitem{additional} We have also added the $U$-independent
eigenvalues \linebreak
$\varepsilon=\pm4$ provided
by the space-antisymmetric states.
\bibitem{rem2} The dimension of this subspace is different from the
one generated by the
$|\varphi_1\rangle$ since, for any $(\alpha, \alpha',K)$ one has~:
$|\varphi_0 (\alpha,\alpha',K)\rangle=|\varphi_0 (\alpha',\alpha,K)\rangle$.
\bibitem{rem1} Such a situation occurs as soon as the two
one-particle wave functions overlap
each other. For example, one can considers a initial configuration
where the two particles are localized on the same site.
\end{thebibliography}
\end{document}